\documentclass{article}

\usepackage{fullpage}
\usepackage{graphicx}
\usepackage{subfigure}
\usepackage{amsmath} 
\usepackage{amsthm}
\usepackage{amsfonts}
\usepackage{amssymb} 
\usepackage{stmaryrd}
\usepackage{bbm} 

\newtheorem{theorem}{Theorem}
\newtheorem{proposition}{Proposition}
\newtheorem{example}{Example}
\newtheorem{remark}{Remark}

\newcommand{\R}{\mathbb{R}} 
\newcommand{\I}{\mathbb{I}} 
\newcommand{\s}{\mathbb{S}} 
\newcommand{\F}{\mathbb{F}} 

\newcommand{\fps}{{\textsf{FPS~}}} 

\newcommand{\rnd}{{\mathrm{fl}}} 
\newcommand{\re}{{\mu}} 

\newcommand{\fun}[1]{{\mathsf{#1}}} 
\newcommand{\ifun}[1]{{[\fun{#1}]}}
\newcommand{\ifunp}[2]{{[\fun{#1}#2]}}

\newcommand{\slope}[3]{{[\mathtt{#1}^{#3}](#2)}} 

\newcommand{\scalar}[1]{{#1}}
\newcommand{\vscalar}[1]{{\vec{#1}}}
\newcommand{\interval}[1]{{\mathrm{#1}}}
\newcommand{\vinterval}[1]{{\vec{\interval{#1}}}}

\newcommand{\varind}{{\mathcal{V}^{\text{ind}}}}
\newcommand{\vardep}{{\mathcal{V}^{\text{dep}}}}

\newcommand{\sem}[1]{\left\llbracket #1 \right\rrbracket_\s^\sharp}
\newcommand{\osem}[1]{\left\llbracket #1 \right\rrbracket^\sharp}

\newcommand{\ie}{\emph{i.e.}~}
\newcommand{\eg}{\emph{e.g.}~}

\title{Interval Slopes as a Numerical Abstract Domain
  \\
  for Floating-Point Variables}

\author {Alexandre Chapoutot}

\author {Alexandre Chapoutot\\
  LIP6 - Universit\'e Pierre et Marie Curie\\
  4, place Jussieur F-75252 Paris Cedex 05 France\\
  \texttt{alexandre.chapoutot@lip6.fr}}

\begin{document}

\maketitle

\begin{abstract}
  The design of embedded control systems is mainly done with model-based
tools such as Matlab/Simu\-link. Numerical simulation is the central
technique of development and verification of such
tools. Floating-point arithmetic, which is well-known to only provide
approximated results, is omnipresent in this activity. In order to
validate the behaviors of numerical simulations using abstract
interpretation-based static analysis, we present, theoretically and
with experiments, a new partially relational abstract domain dedicated
to floating-point variables. It comes from interval expansion of
non-linear functions using slopes and it is able to mimic all the
behaviors of the floating-point arithmetic. Hence it is adapted to
prove the absence of run-time errors or to analyze the numerical
precision of embedded control systems.

\end{abstract}

\section{Introduction}
\label{sec:introduction}

Embedded control systems are made of a software and a physical
environment which aim at continuously interact with each other. The
design of such systems is usually realized with the model-based
paradigm. Matlab/Simulink\footnote{Trademarks of The
  Mathworks\texttrademark company.} is one of the most used tools for
this purpose. It offers a convenient way to describe the software and
the physical environment in an unified formalism. In order to verify
that the control law, implemented in the software, fits the
specification of the system, several numerical simulations are made
under Matlab/Simulink. Nevertheless, this method is closer to
test-based method than formal proof. Moreover, this verification
method is strongly related to the floating-point arithmetic which
provides approximated results.

Our goal is the use of abstract interpretation-based static analysis
\cite{CC77} to validate the design of control embedded software
described in Matlab/Simulink. In our previous work \cite{CM09b}, we
defined an analysis to validate that the behaviors given by numerical
simulations are close to the exact mathematical behaviors. It was
based on an interval abstraction of floating-point numbers which may
produce too coarse results. In this article, our work is focused on a
tight representation of the behaviors of the floating-point arithmetic
in order to increase the precision of the analysis of Matlab/Simulink
models.

To emphasize the poor mathematical properties of the floating-point
arithmetic, let us consider the sum of numbers given in
Example~\ref{exple:ill-conditioned-sums} with a single precision
floating-point arithmetic. The result of this sum is
$-2.08616257.10^{-6}$ due to rounding errors, whereas the exact
mathematical result is zero.
\begin{example}
  \label{exple:ill-conditioned-sums}
  {\small
    \begin{multline*}
      \label{eq:sum-rounding}
      0.0007 + (-0.0097) + 0.0738 + (-0.3122) + 0.7102 + (-0.5709) +
      (-1.0953)\\+ 3.3002 + (-2.9619) + (-0.2353) + 2.4214 +
      (-1.7331) + 0.4121
    \end{multline*}
  }
\end{example}
Example~\ref{exple:ill-conditioned-sums} shows that the summation of
floating-point numbers is a very ill-conditioned problem
\cite[Chap.~6]{handbookfloat09}. Indeed, small perturbations on the
elements to sum produce a floating-point result which could be far
from the exact result. Nevertheless, it is a very common operation in
control embedded software. In particular, it is used in filtering
algorithms or in regulation processes, such as for example in
PID\footnote{PID stands for proportional-integral-derivative. It is a
  generic method of feedback loop control widely used in industry.}
regulation. Remark that depending on the case, the rounding errors may
stay insignificant and the behaviors of floating-point arithmetic may
be safe. In consequence, a semantic model of this arithmetic could be
used to prove the behaviors of embedded control software using
floating-point numbers.

The definition of abstract numerical domains for floating-point
numbers is usually based on rational or real numbers
\cite{Gou01,Min04} to cope with the poor mathematical structure of the
floating-point set. In consequence, these domains give an
over-approximation of the floating-point behaviors. This is because
they do not bring information about the kind of numerical instability
appearing during computations. We underline that our goal is not
interested in computing the rounding errors but the floating-point
result. In others words, we want to compute the bounds of
floating-point variables without considering the numerical quality of
these bounds.

Our main contribution is the definition of a new numerical abstract
domain, called Floating-Point Slopes (\textsf{FPS}), dedicated to the
study of floating-point numbers. It is based on interval expansion of
non-linear functions named \textit{interval slopes} introduced by
Krawczyk and Neumaier \cite{KN85} and, as we will show in this
article, it is a partially relational domain. The main difference is
that, in Proposition~\ref{prop:sound-slope-round-nearest}, we adapt
the interval slopes to deal with floating-point numbers. Moreover, we
are able to tightly represent the behaviors of floating-point
arithmetic with our domain. A few cases studies will show the
practical use of our domain. Hence we can prove properties on programs
taking into account the behaviors of the floating-point arithmetic
such that the absence of run-time errors or, by combining it with
other domains \eg \cite{CM09a}, the quality of numerical computations.

\paragraph{Content.}
In Section~\ref{sec:background}, we will present the main features of
floating-point arithmetic and we will also introduce the interval
expansions of functions. We will present our abstract domain \fps in
Section~\ref{sec:floating-point-slopes} and the analysis of
floating-point programs in
Section~\ref{sec:analysis-of-floating-point-programs} before
describing experimental results in Section~\ref{sec:case-studies}. In
Section~\ref{sec:related-work}, we will reference the related work
before concluding in Section~\ref{sec:conclusion}.

\section{Background}
\label{sec:background}

We recall the main features of the IEEE754-2008 standard of
floating-point arithmetic in
Section~\ref{sec:floating-point-arithmetic}. Next in
Section~\ref{sec:interval-arithmetic}, we present some results from
interval analysis, in particular the interval expansion of functions.

\subsection{Floating-Point Arithmetic}
\label{sec:floating-point-arithmetic}

We briefly present the floating-point arithmetic, more details are
available in \cite{handbookfloat09} and the references therein. The
IEEE754-2008 standard \cite{IEEE754-2008} defines the floating-point
arithmetic in base $2$ which is used in almost every
computer\footnote{It also defines this arithmetic in base $10$ but it
  is not relevant for our purpose.}.

Floating-point numbers have the following form: $f = s . m .  2^e$.
The value $s$ represents the \textit{sign}, the value $m$ is the
\textit{significand} represented with $p$ bits and the value $e$ is
the \textit{exponent} of the floating-point number $f$ which belongs
into the interval $[e_{\min}, e_{\max}]$ such that
$e_{\max}=-e_{\min}+1$. There are two kinds of numbers in this
representation. \textit{Normalize numbers} for which the significand
implicitly starts with a $1$ and \textit{denormalized numbers} that
implicitly starts with a $0$. The later are used to gain accuracy
around zero by slowly degrading the precision.

The standard defines different values of $p$ and $e_{\min}$: $p=24$
and $e_{\min}=-126$ for the single precision and $p=53$ and
$e_{\min}=-1022$ for the double precision. We call \textit{normal
  range} the set of absolute real values in $[2^{e_{\min}},
(2-2^{1-p})2^{e_{\max}}]$ and the \textit{subnormal range} the set of
numbers in $[0, 2^{e_{\min}}[$.

The set of floating-point numbers (single or double precision) is
represented by $\F$ which is closed under negation. A few special
values represent special cases: the values $-\infty$ and $+\infty$ to
represent the negative or the positive overflow; and the value
\textit{NaN}\footnote{\textit{NaN} stands for \textit{Not A Number}.}
represents invalid results such that $\sqrt{-1}$.

The standard defines round-off functions which convert exact real
numbers into floating-point numbers. We are mainly concerned by the
rounding to the nearest ties to even\footnote{The IEEE754-2008
  standard introduces two rounding modes to the nearest with respect
  to the previous IEEE754-1985 and IEEE754-1987 standards. These two
  modes only differ when an exact result is in half-way of two
  floating-point numbers. In rounding-to-nearest-tie-to-even mode, the
  floating-point number whose the least significand bit is even is
  chosen. Note that this definition is used in all the other revisions
  of the IEEE754 standard, see \cite[Chap.~3.4]{handbookfloat09} for
  more details.} (noted $\rnd$), the rounding towards $+\infty$ and
rounding toward $-\infty$. The round-off functions follow the correct
rounding property, \ie the result of a floating-point operation is the
same that the rounding of the exact mathematical result. Note that
these functions are monotone. We are interested in this article by
computing the range of floating-point variables rounded to the nearest
which is the default mode of rounding in computers.

A property of the round-off function $\rnd$ is given in
Equation~\eqref{eq:zero-overflow}. It characterizes the
\textit{overflow}, \ie the rounding result is greater than the biggest
element of $\F$ and the case of the generation of $0$. This definition
only uses positive numbers, using the symmetry property of $\F$, we
can easily deduce the definition for the negative part. We denote by
$\sigma=2^{e_{\text{min}}- p+1}$ the smallest positive subnormal
number and the largest finite floating-point number by
$\Sigma=(2-2^{1-p})2^{e_\text{max}}$. {\small
  \begin{equation}
    \label{eq:zero-overflow}
    \forall x \in\F, x > 0,\quad \rnd(x)=
    \begin{cases}
      +0 & \text{if } 0 < x \leq \sigma/2
      \\
      +\infty & \text{if } x \geq \Sigma
    \end{cases}
  \end{equation}
}An \textit{underflow} \cite[Sect.~2.3]{handbookfloat09} is detected
when the rounding result is less than $2^{e_{\min}}$, \ie the result
is in the subnormal range.

The errors associated to a correct rounding is defined in
Equation~\eqref{eq:rounding-and-error-analysis} and it is valid for
all floating-point numbers $x$ and $y$ except $-\infty$ and $+\infty$
(see \cite[Chap.~2, Sect.~2.2]{handbookfloat09}). The operation
$\diamond \in \{+,-,\times,\div\}$ but it is also valid for the square
root. The \textit{relative rounding error unit} is denoted by
$\re$. In single precision, $\re=2^{-24}$ and $\sigma=2^{-149}$ and in
double precision, $\re=2^{-53}$ and $\sigma=2^{-1074}$.  {\small
  \begin{equation}
    \label{eq:rounding-and-error-analysis}
    \mathtt{fl}(x \diamond y) = (x \diamond y) (1 + \epsilon_1) + \epsilon_2
    \qquad \text{with } 
    |\epsilon_1| \leq \re \text{ and } |\epsilon_2| \leq \frac{1}{2}\sigma
  \end{equation}
}If $\mathtt{fl}(x \diamond y)$ is in the normal range or if $\diamond
\in \{+,-\}$ then $\epsilon_2$ is equal to zero. If $\mathtt{fl}(x
\diamond y)$ is in the subnormal range then $\epsilon_1$ is equal to
zero.

Numerical instabilities in programs come from the rounding
representation of values and they also came from two problems due to
finite precision:
\begin{description}
\item[Absorption] If $|x| \leq \re |y|$ then it happens that $\rnd(x +
  y) = \rnd(y)$. For example, in single precision, the result of
  $\rnd(10^{4} - 10^{-4})$ is $\rnd(10^4)$. In numerical analysis, the
  solution avoid this phenomenon is to sort the sequence of numbers
  \cite[Chap.~4]{Hig02}. This solution is not applicable when the
  numbers to add are given by a sensor measuring the physical
  environment.
\item[Cancellation] It appears in the subtraction $\rnd(x-y)$ if $(|x
  - y|) \leq \re (|x|+|y|)$ then the relative errors can be arbitrary
  big. Indeed, the rounding errors take usually place in the least
  significant digits of floating-point numbers. These errors may
  become preponderant in the result of a subtraction when the most
  significant digits of two closed numbers cancelled each others. In
  numerical analysis, subtraction of numbers coming from long
  computations are avoided to limit this phenomena. We cannot apply
  this solution in embedded control systems where some results are
  used at different instants of time.
\end{description}

\subsection{Interval Arithmetic}
\label{sec:interval-arithmetic}

We introduce interval arithmetic and in particular, the interval
expansion of functions which is an element of our abstract domain
\textsf{FPS}.

\subsubsection{Standard Interval Arithmetic.}
\label{sec:standard-interval-arithmetic}

The \textit{interval arithmetic} \cite{Moo66} has been defined to
avoid the problem of approximated results coming from the
floating-point arithmetic. It had also been used as the first
numerical abstract domain in \cite{CC77}.

When dealing with floating-point intervals the bounds have to be
rounded to outward as in \cite[Sect.~3]{Min04}.  In
Example~\ref{exple:sum-absorption-interval}, we give the result of the
interval evaluation in single precision of a sum of floating-point
numbers.

\begin{example}
  \label{exple:sum-absorption-interval}
  Using the interval domain for floating-point arithmetic
  \cite[Sect.~3]{Min04} the result of the sum defined by
  $\sum_{i=1}^{10}10^1 + \sum_{i=1}^{10}10^2 + \sum_{i=1}^{10}10^3 +
  \sum_{i=1}^{1000}10^{-3}$ is $[11100,11101.953]$. The exact result
  is $11101$ while the floating-point result is $11100$ due to an
  absorption phenomena. The floating-point result and the exact result
  are in the result interval but we cannot distinguish them any more.
\end{example}

A source of over-approximation is known in the interval arithmetic as
the \textit{dependency problem} which is also known in static analysis
as the non-relational aspect. For example, if some variable has value
$[a,b]$, then the result of $x-x$ is $[a-b,b-a]$ which is equal to
zero only if $a=b$. This problem is addressed by
considering interval expansions of functions.

\paragraph{Notations.}
We denote by $\scalar{x}$ a real number and by $\vscalar{x}$ a vector
of real numbers. Interval values are in capital letters $\interval{X}$
or denoted by $[a,b]$ where $a$ is the lower bound and $b$ is the
upper bound of the interval. A vector of interval values will be
denoted by $\vinterval{X}$. We denote by $\ifun{f}$ the interval
extension of a function $\fun{f}$ obtained by substitution of all the
arithmetic operations with their equivalent in interval. The center of
an interval $[a,b]$ is represented by $\fun{mid}([a,b])=a + 0.5 \times
(b - a)$.

\subsubsection{Extended Interval Arithmetic.}
\label{sec:extended-interval-arithmetic}

We are interested in the computation of the image of a vector of
interval $\vinterval{X}$ by a non-linear function $\fun{f}: \R^n
\rightarrow \R$ only composed by additions, subtractions,
multiplications and divisions and square root. In order to reduce
over-approximations in the interval arithmetic, some interval
expansions have been developed. The first one is based on the
\textit{Mean-Value Theorem} and it is expressed as: {\small
  \begin{equation}
    \label{eq:centered-form-of-intervals}
    \fun{f}(\vinterval{X}) \subseteq \fun{f}(\vscalar{z}) + 
    [\fun{f}'](\vinterval{X}) (\vinterval{X} - \vscalar{z})
    \quad \forall \vscalar{z} \in \vinterval{X}
    \enspace .
  \end{equation}
}The first-order approximation of the range of a function $\fun{f}$
can be defined thanks to its first order derivative $\fun{f}'$ over
$\vinterval{X}$. We can then approximate $\fun{f}(\vinterval{X})$ by a
pair $(\fun{f}(\vscalar{z}), \ifunp{f}{'}(\vinterval{X}))$ that are
the value of $\fun{f}$ at point $z$ and the interval extension of
$\fun{f}'$ evaluated over $\vinterval{X}$.

A second interval expansion has been defined by Krawczyk and
Neumaier~\cite{KN85} using the notion of slopes which reduced the
approximation of the derivative form. It is defined by the
relation:{\small
  \begin{equation}
    \label{eq:slope}
    \begin{split}
      \fun{f}(\vinterval{X}) \subseteq \fun{f}(\vscalar{z}) +
      \slope{F}{\vinterval{X}}{\vscalar{z}} (\vinterval{X}-\vscalar{z}) &
      \\
      \qquad \text{with } &
      \mathtt{F}^{\vscalar{z}}(\vinterval{X}) = \left\{
        \frac{\fun{f}(\vscalar{x}) -
          \fun{f}(\vscalar{z})}{\vscalar{x}-\vscalar{z}} : \vscalar{x}
        \in \vinterval{X} \wedge \vscalar{z} \neq \vscalar{x} \right\} 
      \enspace .
    \end{split}
  \end{equation}
}Then we can represent $\fun{f}(\vinterval{X})$ by a pair
$(\fun{f}(\vscalar{z}), \slope{F}{\vinterval{X}}{\vscalar{z}})$ that
are the value of $\fun{f}$ in the point $\vscalar{z}$ and the interval
extension of the slope $\mathtt{F}^z(X)$ of $\fun{f}$.

Note that the value $\vscalar{z}$ is constructed, in general, from the
centers of the interval variables appearing in the function $\fun{f}$
for both interval expansions.

An interesting feature is that we can inductively compute the
derivative or the slope of a functions using \textit{automatic
  differentiation} techniques \cite{BHN02}. It is a semantic-based
method to compute derivatives. In this context, we call
\textit{independent variables} some input variables of a program with
respect to which derivatives are computed. We call \textit{dependent
  variables} output variables whose derivatives are desired. A
\textit{derivative object} represents derivative information, such as
a vector of partial derivatives like $(\partial e/\partial x_1,
\dots, \partial e/\partial x_n)$ of some expression $e$ with respect
to a vector $\vscalar{x}$ of independent variables. The main idea of
automatic differentiation is that every complicated function
$\fun{f}$, \ie a program, is composed by simplest elements, \ie
program instructions. Knowing the derivatives of these elements with
respect to some independent variables, we can compute the derivatives
or the slopes of $\fun{f}$ following the differential calculus
rules. Furthermore, using interval arithmetic in the differential
calculus rules, we can guarantee the result.

We give in Table~\ref{tab:automatic-differentiation-rules} the rules
to compute derivatives or slopes with respect to the structure of
arithmetic expressions. We assume that we know the number of
independent variables in the programs and we denote by $n$ this
number. The variable $\varind$ represents the vector of independent
variables with respect to which the derivatives are computed. We
denote by $\delta_i$ the interval vector of length $n$, having all its
coordinates equal to $[0,0]$ except the $i$-th element equals to
$[1,1]$. So, we consider that all the independent variables are
assigned to a unique position $i$ in $\varind$ and it is initially
assigned with a derivative object equal to $\delta_i$. Following
Table~\ref{tab:automatic-differentiation-rules} where $\fun{g}$ and
$\fun{h}$ represent variables with derivative object, a constant value
$c$ has a derivative object equal to zero (the interval vector
$\vinterval{0}$ has all its coordinates equal to $[0,0]$). For
addition and subtraction, the result is the vector addition or the
vector subtraction of the derivative objects. For multiplication and
division, it is more complicated but the rules come from the standard
rules of the composition of derivatives, \eg $(u\times v)'=u'\times v
+ u \times v'$. A proof of the computation rules\footnote{In
  \cite[Sect.~2]{KN85}, the authors went also into detail of the
  complexity of these operations.} for slopes can be found in
\cite[Sect.~1]{Rump96}. Note that we can apply automatic
differentiation for other functions, such as the square root, using
the rule of function composition, $(f \circ g)'(x)=f'(g(x))g'(x)$.

These interval expansions of functions, using either
$(\fun{f}(\vscalar{z}), \ifunp{f}{'}(\vinterval{X}))$ the derivative
form or $(\fun{f}(\vscalar{z}),
\slope{f}{\vinterval{X}}{\vscalar{z}})$ the slope form, define a
straightforward semantics of arithmetic expressions which can be used
to compute bounds of variables.

\begin{table}[htbp]
  \centering
  \caption{Automatic differentiation rules for derivatives and slopes}
  \label{tab:automatic-differentiation-rules}
  \begin{tabular}{c c p{1em} c}
    \hline\noalign{\smallskip} 
    Function & Derivative arithmetic & & Slope arithmetic
    \\
    \noalign{\smallskip} 
    \hline 
    \noalign{\smallskip} 
    $c \in \R$ & $\mathbf{0}$ && $\mathbf{0}$
    \\
    \begin{math}
      \fun{g} + \fun{h}
    \end{math}
    & 
    \begin{math}
      [\fun{g}'](\vinterval{X})+[\fun{h}'](\vinterval{X})
    \end{math}
    &&
    \begin{math}
      \slope{G}{\vinterval{X}}{\vscalar{z}} + 
      \slope{H}{\vinterval{X}}{\vscalar{z}}
    \end{math}
    \\
    \begin{math}
      \fun{g} - \fun{h}
    \end{math}
    &
    \begin{math}
      [\fun{g}'](\vinterval{X})- [\fun{h}'](\vinterval{X})
    \end{math}
    &&
    \begin{math}
      \slope{G}{\vinterval{X}}{\vscalar{z}} - 
      \slope{H}{\vinterval{X}}{\vscalar{z}}
    \end{math}
    \\
    \begin{math}
      \fun{g} \times \fun{h}
    \end{math}
    &
    \begin{math}
      [\fun{g}'](\vinterval{X}) \times \fun{h}(\vinterval{X}) + 
      \fun{g}(\vinterval{X}) \times [\fun{h}'](\vinterval{X})
    \end{math}
    &&
    \begin{math}
      \slope{G}{\vinterval{X}}{\vscalar{z}} \times \fun{h}(\vinterval{X}) + 
      \fun{g}(\vscalar{z}) \times \slope{H}{\vinterval{X}}{\vscalar{z}}
    \end{math}
    \\
    \begin{math}
      \displaystyle \frac{\fun{g}}{\fun{h}}
    \end{math}
    &
    \begin{math}
     \displaystyle \frac{[\fun{g}'](\vinterval{X})\times \fun{h}(\vinterval{X})
       - [\fun{h}'](\vinterval{X})\times \fun{g}(\vinterval{X})}
     {\fun{h}^2(\vinterval{X})}
    \end{math}
    &&
    \begin{math}
      \displaystyle 
      \frac{\slope{G}{\vinterval{X}}{\vscalar{z}} - 
        \slope{H}{\vinterval{X}}{\vscalar{z}}\times 
        \frac{\fun{g}(\vscalar{z})}{\fun{h}(\vscalar{z})}}
      {\fun{h}(\vinterval{X})}
    \end{math}
    \\
    \begin{math}
      \displaystyle \sqrt{\fun{g}}
    \end{math}
    & 
    \begin{math}
      \displaystyle 
      \frac{1}{2}\frac{[\fun{g}'](\vinterval{X})}{\sqrt{\fun{g}(\vinterval{X})}}
    \end{math}
    &&
    \begin{math}
      \displaystyle 
      \frac{\slope{G}{\vinterval{X}}{\vscalar{z}}}{\sqrt{g(\vscalar{z})}
        +\sqrt{g(\vinterval{X})}}
    \end{math}
    \\
    \hline 
  \end{tabular}
\end{table}

\begin{remark}
  The difference in over-approximated result between the derivative
  form and the slope form is in the multiplication and the division
  rules. In the derivative form, we need to evaluate the two operands
  ($\fun{g}$ and $\fun{h}$) using interval arithmetic while we only
  need to evaluate one of them in the slope form. Note also that we
  could have defined the multiplication by
  $\slope{H}{\vinterval{X}}{\vscalar{z}} \times \fun{g}(\vinterval{X})
  + \fun{h}(\vscalar{z}) \times \slope{G}{\vinterval{X}}{\vscalar{z}}$
  (the division has also two forms) but the two possible forms of
  slope are over-approximations of
  $\fun{f}(\vscalar{X})$. Nevertheless, a possible way to choose
  between the two forms is to keep the form which gives the smallest
  approximation of $\fun{f}(\vscalar{X})$.
\end{remark}

\begin{figure}[ht]
  \centering
  \subfigure[{$x\in[-1,1/2]$}]{
    \label{fig:graphical-example-slope-large-interval}
    \includegraphics[scale=0.2]{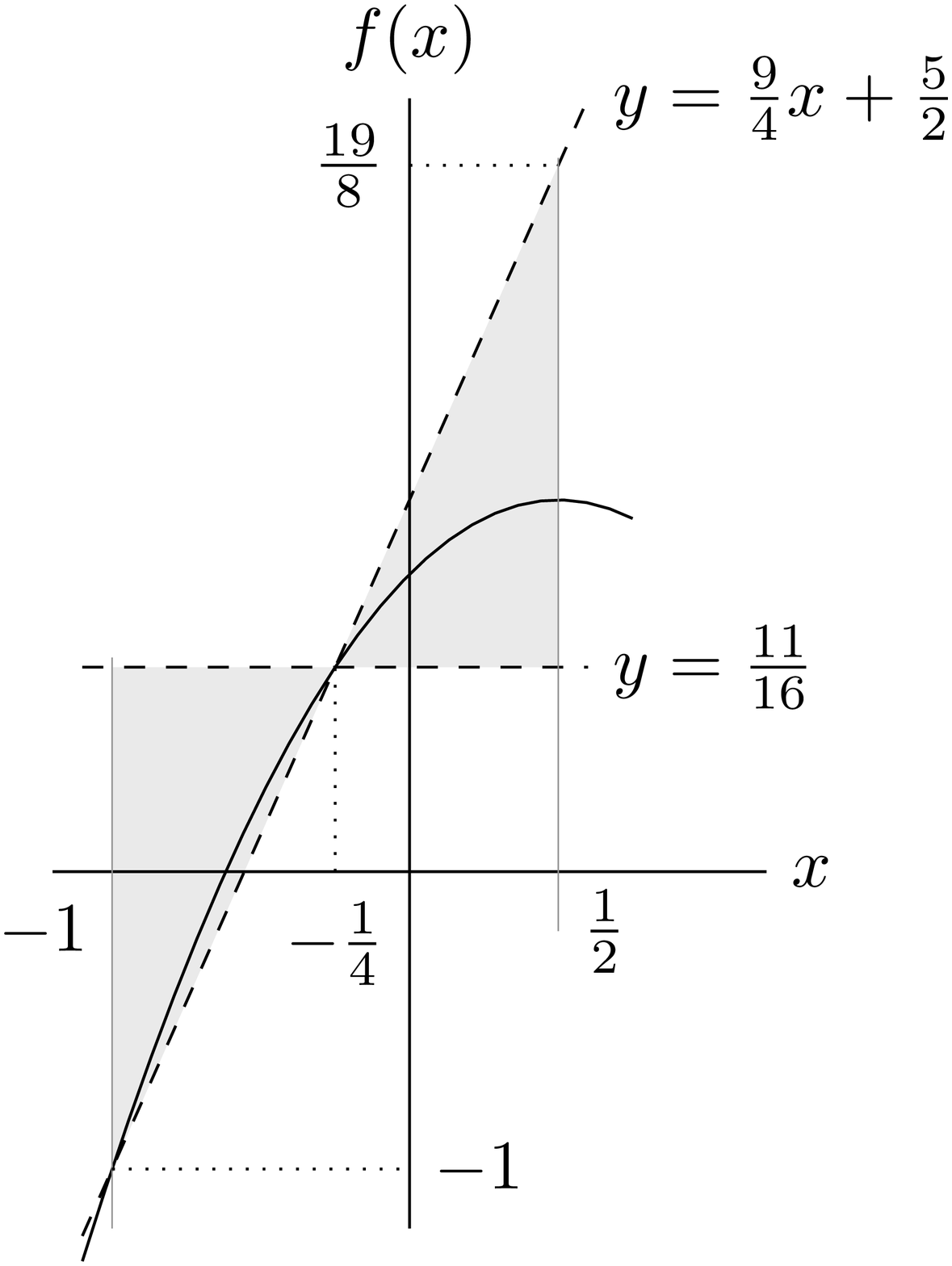}
  }
  \qquad
  \subfigure[{$x\in[-1/2,1/2]$}]{
    \label{fig:graphical-example-slope-small-interval}
    \includegraphics[scale=0.2]{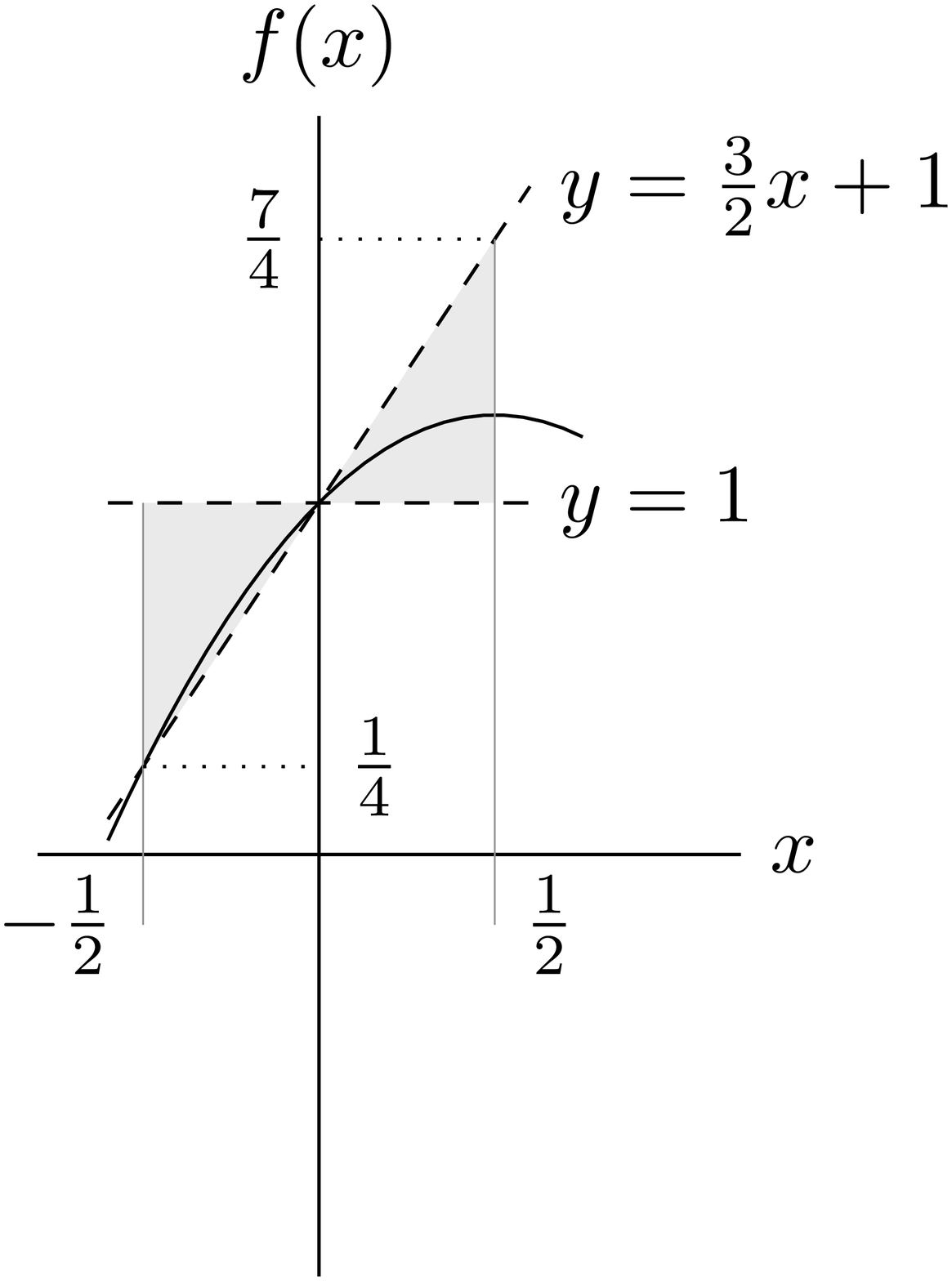}
  }
  \caption{Two examples of the interval expansion with slopes}
\label{fig:graphical-example-slope}
\end{figure}

In Figure~\ref{fig:graphical-example-slope}, we give two graphical
representations of interval slope expansion. For this purpose, we want
to compute the image of $x$ by the function $f(x)=x(1-x)+1$. We
consider in Figure~\ref{fig:graphical-example-slope-large-interval}
that $x\in[-1,1/2]$ and we get as a result that $f(x)\in[-1,19/8]$
which is an over-approximation of the exact result $[-1,5/4]$. The
midpoint is $-1/4$ and the set of slopes is bounded by the interval
$[0,9/4]$. The dashed lines represent the linear approximation of the
image. In Figure~\ref{fig:graphical-example-slope-small-interval}, we
consider that $x\in[-1/2, 1/2]$ and the result is $f(x)\in [1/4, 7/4]$
which is still an over-approximation of the exact result $[1/4,
5/4]$. In that case, the midpoint is $0$ and the set of slopes is
bounded by the interval $[0,3/2]$. Note that the smaller the interval
the better the approximation is.

Example~\ref{ex:computing-slopes} shows that we can encode with
interval slopes the list of variables contributing in the result of an
arithmetic expression. In particular, the vector composing the
interval slope of the variable $t$ represents the influence of the
variables $a$, $b$ and $c$ on the value of $t$. For example, we know
that a modification of the value of the variable $a$ produce a
modification of the result with the same order of the modification on
$a$ because the slope associated to $a$ is $[1,1]$. But a modification
on the variable $b$ by $\Delta_b$ will produce a modification on the
$t$ by $\Delta_b \times \interval{V}_c$ because the slope of $b$ is
equal to $\interval{V}_c$.

\begin{example}
  \label{ex:computing-slopes}
  Let $t = a + b \times c$, we want to compute the interval slope
  $\slope{T}{\vinterval{X}}{z}$ of $t$. We consider that
  $\varind=\{a,b,c\}$ and $\vinterval{\vinterval{X}}$ is the interval
  vector of the values of these variables. We suppose that the
  interval slope expansion of $a$, $b$ and $c$ are $(z_a,
  \slope{A}{\vinterval{X}}{\vscalar{z}} = \delta_1)$, $(z_b,
  \slope{B}{\vinterval{X}}{\vscalar{z}} = \delta_2)$, and $(z_c,
  \slope{C}{\vinterval{X}}{\vscalar{z}} = \delta_3)$ respectively. The interval
  value associated to $c$ is $\interval{V}_c$ \ie $\interval{V}_c =
  z_c + \slope{C}{\vinterval{X}}{z}(\vinterval{X}-\vscalar{z})$. {\small
    \begin{align*}
      \slope{T}{\vinterval{X}}{\vscalar{z}} & = 
      \slope{A}{\vinterval{X}}{\vscalar{z}} +
      z_b \slope{C}{\vinterval{X}}{\vscalar{z}} +
      \slope{B}{\vinterval{X}}{\vscalar{z}}
      \left(z_c+\slope{C}{\vinterval{X}}{\vscalar{z}}
        (\vinterval{X}-\vscalar{z})
      \right)
      \\
      & = ([1,1], 0, 0) + z_b \times (0, 0, [1,1]) + (0, [1,1], 0)
      \times \interval{V}_c
      \\
      & = \left([1,1], [1,1] \times \interval{V}_c, z_b \times
        [1,1]\right)
      \\
      & = \left( [1,1], \interval{V}_c, [z_b,z_b]\right)
    \end{align*}}
\end{example}

As seen in Example~\ref{ex:computing-slopes}, interval slopes
represent relations between the inputs and the outputs of a
function. By computing interval slopes, we build step by step the set
of variables related to arithmetic expressions in programs. In static
analysis, we can use this interval expansion to track the influence of
the inputs of a program on its outputs. Hence the choice of the set
$\varind$ of independent variables is given by the set of the input
variables of the program to analyse. Moreover, we can add in $\varind$
all the other variables which may influence
output.

\section{Floating-Point Slopes}
\label{sec:floating-point-slopes}

We present in this section our new abstract domain \texttt{FPS}. In
Section~\ref{sec:floating-point-version-of-interval-slopes}, we adapt
the computation rules of interval slopes to take into account
floating-point arithmetic. Next in
Section~\ref{sec:arithmetic-operations}, we define an abstract
semantics of arithmetic expressions over \texttt{FPS} values taking
into account the behaviors of floating-point arithmetic. And in
Section~\ref{sec:order-structure}, we define the order structure of
the \texttt{FPS} domain.

\subsection{Floating-Point Version of Interval Slopes}
\label{sec:floating-point-version-of-interval-slopes}

The definition of interval slope expansion in
Section~\ref{sec:interval-arithmetic} manipulates real numbers. In
case of floating-point numbers, we have to take into account the
round-off function and the rounding-errors.

We show in Proposition~\ref{prop:sound-slope-round-nearest} that the
range of a non-linear function $\fun{f}$ of floating-point numbers can
be soundly over-approximated by a floating-point slope. The function
$\fun{f}$ must respect the correct rounding, \ie the property of
Equation~\eqref{eq:rounding-and-error-analysis} must hold. In other
words, the result of an operation over set of floating-point numbers
is over-approximated by the result of the same operation over
floating-point slopes by adding a small quantity depending on the
relative rounding error unit $\re$ and the absolute error $\sigma$.

\begin{proposition}
  \label{prop:sound-slope-round-nearest}
  Let $\fun{f}: D \subseteq \R^n \rightarrow \R$ be an arithmetic
  operation of the form $g \diamond h$ with $\diamond \in
  \{+,-,\times,\div\}$ or $\sqrt{~}$, \ie $\fun{f}$ respects the
  correct rounding. For all $\vinterval{X} \subseteq D$ and
  $\vscalar{z} \in D$, we have: {\small
    \begin{displaymath}
      \rnd\big(\fun{f}(\vinterval{X})\big) 
      \subseteq \fun{f}(\vscalar{z})\big
      (1 + [-\re,\re]\big) + 
      \left[-\frac{\sigma}{2}, \frac{\sigma}{2}\right] +
      \slope{F}{\vinterval{X}}{\vscalar{z}}(X-z)\big(1+[-\re,\re]\big)
      \enspace .
    \end{displaymath}
  }
  \begin{proof}
    {\small  \begin{align*}
        \rnd\big(\fun{f}(\vinterval{X})\big) 
        & = \{ \fun{f}(\vscalar{x}) (1+\varepsilon_x) + 
        \bar{\varepsilon}_x : \vscalar{x} \in \vinterval{X}\}
        && \text{by Eq.~\eqref{eq:rounding-and-error-analysis}}
        \\
        & \subseteq
        \fun{f}(\vinterval{X}) + 
        \fun{f}(\vinterval{X})\{ \varepsilon_x : \vscalar{x} 
        \in \vinterval{X}\} 
        + \{ \bar{\varepsilon}_x : \vscalar{x} \in \vinterval{X}\}
        &&
        \\
        & \subseteq
        \big(\fun{f}(\vscalar{z}) + \slope{F}{\vinterval{X}}{\vscalar{z}}
        (\vinterval{X}-\vscalar{z})\big) + 
        \{ \bar{\varepsilon}_x : \vscalar{x} \in \vinterval{X}\}
        &&  \text{by Eq.~\eqref{eq:slope}} 
        \\
        & \quad + \big(\fun{f}(\vscalar{z}) + 
        \slope{F}{\vinterval{X}}{\vscalar{z}}(\vinterval{X}-\vscalar{z})\big) 
        \{ \varepsilon_x : \vscalar{x} \in \vinterval{X}\} 
        &&
        \\
        & \subseteq \fun{f}(\vscalar{z}) 
        \big(1+\{ \varepsilon_x : \vscalar{x} \in \vinterval{X}\}\big) +
        \{ \bar{\varepsilon}_x : \vscalar{x} \in \vinterval{X}\}
        &&
        \\
        &
        \quad + \slope{F}{\vinterval{X}}{\vscalar{z}}
        (\vinterval{X}-\vscalar{z}) 
        \big(1+\{ \varepsilon_x : \vscalar{x} \in \vinterval{X}\}\big)
        &&
        \\
        & \subseteq 
        \fun{f}(\vscalar{z}) \big(1+[-\re,\re]\big) + 
        \left[-\frac{\sigma}{2},\frac{\sigma}{2}\right]
        && |\varepsilon_x| \leq \re
        \text{ by Eq.~\eqref{eq:rounding-and-error-analysis}}
        \\
        & \quad + \slope{F}{\vinterval{X}}{\vscalar{z}}
        (\vinterval{X}-\vscalar{z}) 
        \big(1+[-\re,\re]\big)
        &&
        |\bar{\varepsilon}_x| \leq \frac{1}{2}\sigma
        \text{ by Eq.~\eqref{eq:rounding-and-error-analysis}}
      \end{align*}
      \qed
    }
  \end{proof}
\end{proposition}

\begin{remark}
  As the floating-point version of slopes is based on $\re$ and
  $\sigma$, we can represent the floating-point behaviors depending of
  the hardware. For example, extended precision\footnote{In some
    hardware, \eg Intel x87, floating-point numbers may be encoded
    with $80$ bits in registers, \ie the significand is $64$ bits
    long.} is represented using the values $\re=2^{-64}$ and
  $\sigma=2^{-16446}$. Furthermore following \cite{BN10}, we can
  compute the result of a double rounding\footnote{It may happen on
    hardware using extended precision. Results of computations are
    rounded in registers and they are rounded again, with a less
    precision, in memory.}  with $\re=(2^{11}+2)2^{-64}$ and
  $\sigma=(2^{11}+1)2^{-1086}$. 
\end{remark}

Proposition~\ref{prop:sound-slope-round-nearest} shows that we can
compute the floating-point range of a function $\fun{f}$, respecting
the correct rounding, using interval slopes expansion. That is a set
of floating-point values can is represented by a pair:{\small
  \begin{displaymath}
    \left(
      \left[\fun{f}\right](\vinterval{z})
      \big(1 + \left[-\re,\re\right]\big) + 
      \left[-\frac{\sigma}{2}, \frac{\sigma}{2}\right],
      \quad
      \slope{F}{\vscalar{X}}{\vscalar{z}}\big(1+[-\re,\re]\big)
    \right)
    \enspace.
  \end{displaymath}
}The first element is a small interval rounding to the nearest around
$\fun{f}(\vscalar{z})$ for which we have to take into account the
possible rounding errors. The second element is the interval slopes
which have to take account of relative errors. Note that this
adaptation adds a very little overhead of computations compared to the
definition of interval slopes by Krawczyk and Neumaier.

\subsection{Semantics of Arithmetic Operations}
\label{sec:arithmetic-operations}

In this section, we define the abstract semantics of arithmetic
operations over elements of floating-point slopes domain in order to
mimic the behaviors of the floating-point arithmetic. We denote by
$\I$ the set of intervals and by $\s = \I \times \I^{|\varind|}$ the
set of slopes. An element $s$ of $\s$ is represented by a pair
$(\interval{M}, \vinterval{S})$ where $\interval{M}$ is a
floating-point interval and $\vinterval{S}$ is a vector of
floating-point intervals. We denote by $\langle \I, \sqsubseteq_\I,
\bot_\I, \top_\I, \sqcup_\I, \sqcap_\I \rangle$ the lattice of
intervals. First we define some auxiliary functions before presenting
the semantics of arithmetic expressions over \textsf{FPS}.

The function $\iota$ defined in Equation~\eqref{eq:iota} computes the
interval value associated to a floating-point slopes $(\interval{M},
\vinterval{S})$. We assume that the values of independent variables
are kept in a separate interval vector $\vinterval{V}_\varind$. The
notation $\fun{mid}(\vinterval{V}_\varind)$ stands for the
component-wise application of the function $\fun{mid}$ on all the
components of the vector $\vinterval{V}_\varind$. Note that $\cdot$
represents the scalar product. {\small
\begin{equation}
  \label{eq:iota}
  \iota\big( (\interval{M}, \vinterval{S}) \big) = 
  \interval{M} + \vinterval{S} \cdot 
  \big(\vinterval{V}_\varind - \fun{mid}(\vinterval{V}_\varind)\big)
\end{equation}
}The function $\kappa$ defined in Equation~\eqref{eq:kappa} transforms
an interval value $[a,b]^\ell$ associated to the $\ell$-th independent
variable into a floating-point slope. {\small
  \begin{equation}
    \label{eq:kappa}
    \kappa\left( [a,b]^\ell\right) = \big( [m,m], \delta_\ell \big)
    \quad\text{with}\quad m=\fun{mid}([a,b])
  \end{equation}
}This function $\kappa$ is used in two cases: \textit{i)}~To
initialize all the independent variables at the beginning of an
analysis. \textit{ii)}~In the meet operation, see
Section~\ref{sec:order-structure}.

We can detect overflows and generations of zero by using the function
$\Phi$ defined in Equation~\eqref{eq:fps-zero-overflow}. We have two
kinds or rules: \textit{total} rules when we are certain that a zero
or an overflow occur and \textit{partial} rules when a part of the set
described by a floating-point slope generates a zero or an
overflow. With the function $\iota$ we can determine for an element
$(\interval{M},\vinterval{S}) \in \s$ if $(\interval{M},
\vinterval{S})$ represents an overflow or a zero. Hence we represent
the finite precision of the floating-point arithmetic. We denote by
$\text{\textbf{p}}_\infty$ and by $\text{\textbf{m}}_\infty$ the
interval vectors with all their components equal to
$[+\infty,+\infty]$ and $[-\infty,-\infty]$ respectively. We recall
that $\sigma$ is the smallest denormalized and $\Sigma$ is the largest
floating-point numbers. {\small
  \begin{equation}
    \label{eq:fps-zero-overflow}
    \Phi(\interval{M}, \vinterval{S}) = 
    \begin{cases}
      ([0,0], \mathbf{0}) 
      & \text{if } \iota(\interval{M}, \vinterval{S}) \sqsubseteq_\I
      [-\frac{\sigma}{2},\frac{\sigma}{2}]
      \\
      (\tilde{\interval{M}},
      \mathbf{0}\ \dot{\sqcup}_\I\ S) 
      & \text{if } \iota(\interval{M}, \vinterval{S})
      \ \sqcap_\I\ ]-\frac{\sigma}{2},\frac{\sigma}{2}[\ \neq \bot_\I
      \\
      & \text{and }\tilde{\interval{M}} =
      \begin{cases}
        [0,0] & \text{if } \interval{M}\ 
        \sqsubseteq_\I\ ]-\frac{\sigma}{2},\frac{\sigma}{2}[
        \\
        [0,0] \sqcup_\I \interval{M} & \text{otherwise}
      \end{cases}
      \\
      ([+\infty, +\infty], \text{\textbf{p}}_\infty) 
      & \text{if } \iota(\interval{M}, \vinterval{S}) \sqsubseteq_\I\ 
      ]\Sigma,+\infty]
      \\
      (\tilde{\interval{M}}, \text{\textbf{p}}_\infty\ \dot{\sqcup}_\I\ S) & 
      \text{if }
      \iota(\interval{M}, \vinterval{S})\ 
      \sqcap_\I\ ]\Sigma,+\infty] \neq \bot_\I
      \\
      & \text{and }\tilde{\interval{M}} =
      \begin{cases}
        [+\infty, +\infty] & \text{if } 
        \interval{M}\ \sqsubseteq_\I\ ]\Sigma,+\infty]
        \\
        [+\infty,+\infty] \sqcup_\I \interval{M} & \text{otherwise}
      \end{cases}
      \\
      ([-\infty, -\infty], \text{\textbf{m}}_\infty) 
      & \text{if } \iota(\interval{M}, \vinterval{S}) \sqsubseteq_\I
      [-\infty, -\Sigma[
      \\
      (\tilde{\interval{M}}, 
      \text{\textbf{m}}_\infty\ \dot{\sqcup}_\I\ S) & \text{if }
      \iota(\interval{M}, \vinterval{S})
      \sqcap_\I [-\infty, -\Sigma[\ \neq \bot_\I
      \\
      & \text{and }\tilde{\interval{M}} =
      \begin{cases}
        [-\infty, -\infty] & \text{if } 
        \interval{M}\ \sqsubseteq_\I\ [-\infty, -\Sigma[
        \\
        [-\infty,-\infty] \sqcup_\I \interval{M} & \text{otherwise}
      \end{cases}
      \\
      (\interval{M}, \vinterval{S}) & \text{otherwise}
    \end{cases}
  \end{equation}
}Equation~\eqref{eq:fps-zero-overflow} is an adaptation of the rule
defined in Equation~\eqref{eq:zero-overflow} to deal with \fps
values. Furthermore, the abstract values $(+\infty,
\mathbf{p}_\infty)$ and $(-\infty, \mathbf{m}_\infty)$ represent the
special floating-point values $+\infty$ and $-\infty$ respectively. As
in floating-point arithmetic, the values $(+\infty,
\mathbf{p}_\infty)$ and $(-\infty, \mathbf{m}_\infty)$ are absorbing
elements.

An interesting feature of interval slopes is that we can mimic the
absorption phenomenon by setting to zero the interval slope of the
absorbed operand. We define the function $\rho$ for this
purpose. Indeed, an abstract value $(\interval{M}, \vinterval{S})$
already supports partial absorption as $\interval{M}$ is computed with
a rounding to the nearest but $\vinterval{S}$ have to be
\textit{reduced} to represent the absence of the influence of
particular independent variables. The reduction of an abstract value
$g=(\interval{M}_g, \vinterval{S}_g)$ compared to an abstract value
$h=(\interval{M}_h, \vinterval{S}_h)$, denoted by $\rho(g \mid h)$, is
defined in Equation~\eqref{eq:slope-reduction}.  {\small
  \begin{equation}
    \label{eq:slope-reduction}
    \rho(g \mid h) =         
    \begin{cases}
      ([0, 0], \mathbf{0}) & \text{if } 
      \iota\big(\interval{M}_g, \vinterval{S}_g\big)
      \sqsubseteq_\I [\re,\re]
      \times \iota\big(\interval{M}_h, \vinterval{S}_h\big)
      \\
      \big(\tilde{\interval{M}}_g, 
      \mathbf{0}\ \dot{\sqcup}_\I\ \vinterval{S}_g\big) 
      & \text{if } \iota
      \big(\interval{M}_g, \vinterval{S}_g\big) \sqcap_\I 
      [\re, \re] \times \iota\big(\interval{M}_h, \vinterval{S}_h\big)
      \neq \bot_\I
      \\
      & \text{and } \tilde{\interval{M}}_g =
      \begin{cases}
        [0, 0] & \text{if } \interval{M}_g\ \sqsubseteq_\I\ [\re,\re]
        \times
        \iota\big(\interval{M}_h, \vinterval{S}_h\big)
        \\
        [0, 0] \sqcup_\I \interval{M}_g & \text{otherwise}
      \end{cases}
      \\
      \big(\interval{M}_g, \vinterval{S}_g\big) & \text{otherwise}
    \end{cases}
  \end{equation}
}Equation~\eqref{eq:slope-reduction} models the absorption phenomenon
by explicitly setting to zero the values of a slope. As mentioned in
Section~\ref{sec:interval-arithmetic}, a slope shows which variables
influence the computation of an arithmetic expression. But, absorption
phenomena induce that an operand does not influence the result of an
addition or a subtraction any more.

Using the functions $\Phi$, $\rho$ and $\iota$, we inductively define
on the structure of arithmetic expressions the abstract semantics
$\sem{.}$ of floating-point slopes in
Figure~\ref{fig:arithmetic-operation-floating-point-slopes}. We denote
by $\text{env}^\sharp$ an abstract environment which associates to
each program variable a floating-point slope. For each arithmetic
operation, we component-wisely combine the elements of the abstract
operands $\sem{g}(\text{env}^\sharp)=(\interval{M}_g,
\vinterval{S}_g)$ and $\sem{h}(\text{env}^\sharp)=(\interval{M}_h,
\vinterval{S}_h)$. The element $\interval{M}$ is obtained using the
interval arithmetic with rounding to the nearest. The element
$\vinterval{S}$ is computed using the definition of the slope
arithmetic defined in Table~\ref{tab:automatic-differentiation-rules}.
We take into account of the possible rounding errors in the result
$(\interval{M}, \vinterval{S})$ following
Proposition~\ref{prop:sound-slope-round-nearest}. In case of addition
and subtraction, according to the
Equation~\eqref{eq:rounding-and-error-analysis}, we do not consider
absolute error $\frac{\sigma}{2}$ which is always zero. Moreover, in
case of addition or subtraction, we handle the absorption phenomena
using the function $\rho$, defined in
Equation~\eqref{eq:slope-reduction}. Finally, we check if a zero or an
overflow is generated by applying the function $\Phi$ defined in
Equation~\eqref{eq:fps-zero-overflow}.

\begin{figure}[ht] {\small
    \begin{displaymath}
      \begin{split}
        \sem{g \pm h}(\theta^\sharp) & =\Phi\left( (\tilde{\interval{M}}_g 
          \pm \tilde{\interval{M}}_h)(1+[-\re,\re]),\quad
          \left(\tilde{\vinterval{S}}_g \pm 
            \tilde{\vinterval{S}}_h\right)(1+[-\re,\re]) \right)
        \\
        &
        \begin{flalign*}
          \text{with}
          \quad (\tilde{\interval{M}}_g,\tilde{\vinterval{S}}_g) = 
          \rho(g \mid h)
          \text{ and } (\tilde{\interval{M}}_h,\tilde{\vinterval{S}}_h)
          = \rho(h \mid g)
        \end{flalign*}
        \\
        \sem{g \times h}(\theta^\sharp) & = \Phi 
        \left(\interval{M},\quad
          \big(
          \vinterval{S}_g \times \iota(\interval{M}_h, \vinterval{S}_h) 
          + \interval{M}_g \times \vinterval{S}_h
          \big) (1+[-\re,\re])        
        \right)
        \\
        &
        \begin{flalign*}
          \text{with}
          \quad \interval{M} = 
          (\interval{M}_g \times \interval{M}_h)(1 + [-\re,\re]) + 
          \left[\frac{\sigma}{2}, \frac{\sigma}{2}\right]
        \end{flalign*}
        \\
        \sem{\frac{g}{h}}(\theta^\sharp) & = 
        \Phi \left(
          \interval{M},\quad
          \frac{\vinterval{S}_g - \vinterval{S}_h
            \frac{\interval{M}_g}{\interval{M}_h}}
          {\iota(\interval{M}_h, \vinterval{S}_h)} (1+[-\re,\re])
        \right) 
        \\
        &
        \begin{flalign*}
          \begin{split}
          \text{with}\quad
          \interval{M} & =  
          \frac{\interval{M}_g}{\interval{M}_h}(1 + [-\re,\re]) + 
          \left[\frac{\sigma}{2}, \frac{\sigma}{2}\right],
          \\
          0 & \not\in \iota(\interval{M}_h, \vinterval{S}_h) 
          \text{ and } 0 \not\in \interval{M}_h
        \end{split}
        \end{flalign*}
        \\              
        \sem{\sqrt{g}}(\theta^\sharp) & = 
        \Phi \left(
        \interval{M},\quad
        \left(
          \frac{\vinterval{S}_g}{\sqrt{\interval{M}_g}+
            \sqrt{\iota(\interval{M}_g, \vinterval{S}_g)}} 
        \right) (1+[-\re,\re])
      \right)
      \\
      &
      \begin{flalign*}
        \begin{split}
        \text{with}\quad 
        \interval{M} &= 
        \left(\sqrt{\interval{M}_g}\left(1+[-\re,\re]\right)\right)+
        \left[-\frac{\sigma}{2},\frac{\sigma}{2}\right],
        \\
        \interval{M}_g & \sqcap_\I [-\infty,0] = \bot_\I        
        \text{ and } \iota(\interval{M}_g, \vinterval{S}_g) 
         \sqcap_\I [-\infty,0] = \bot_\I
      \end{split}
      \end{flalign*}
    \end{split}
  \end{displaymath}
}
\caption{Abstract semantics of arithmetic expressions on
  floating-point slopes}
\label{fig:arithmetic-operation-floating-point-slopes}
\end{figure}

\begin{remark}
  The functions $\Phi$ and $\rho$ make the arithmetic operations on
  floating-point slopes non associative and non distributive as in
  floating-point arithmetic.
\end{remark}

\subsection{Order Structure}
\label{sec:order-structure}

In this section, we define the order structure of the set $\s$ of
floating-point slopes. In particular, this structure is based on the
lattice of intervals. We recall that the set of slopes $\s = \I \times
\I^{|\varind|}$ and an element $s$ of $\s$ is a pair $(\interval{M},
\vinterval{S})$.

We define a partial order, the join and the meet operations between
elements of $\s$. All these operations are defined as a component-wise
application of the associated operations of the interval domain except
the meet operation which needs extra care. We denote by
$\dot{\sqsubseteq}_\I$ the component-wise application of the interval
order. We can define a partial order $\sqsubseteq_\s$ between elements
of $\s$ with: {\small
  \begin{multline}
    \label{eq:slopes-order}
    \forall (\interval{M}_g, \vinterval{S}_g), (\interval{M}_h, \vinterval{S}_h)
    \in \s,
    \left(\interval{M}_g, \vinterval{S}_g\right) 
    \sqsubseteq_\s \left(\interval{M}_h, \vinterval{S}_h\right)
    \Leftrightarrow \interval{M}_g \sqsubseteq_\I \interval{M}_h 
    \wedge \vinterval{S}_g\
    \dot{\sqsubseteq}_\I\ \vinterval{S}_h\enspace .
  \end{multline}
}

The join operation $\sqcup_\s$ over floating-point slopes is defined
in Equation~\eqref{eq:join-operation}. We denote by $\dot{\sqcup}_\I$
the component-wise application of the operation $\sqcup_\I$.  {\small
  \begin{multline}
    \label{eq:join-operation}
    \forall (\interval{M}_g, \vinterval{S}_g), 
    (\interval{M}_h, \vinterval{S}_h) \in \s, \quad
    \big(\interval{M}_g, \vinterval{S}_g\big) \ 
    \sqcup_\s\ \big(\interval{M}_h, \vinterval{S}_h\big) = \big(\interval{M},
    S\big) 
    \\
    \text{with}\quad
    \interval{M} = \interval{M}_g 
    \sqcup_\I \interval{M}_h \quad\text{and}\quad S = \vinterval{S}_g\
    \dot{\sqcup}_\I\ \vinterval{S}_h 
  \end{multline}
}

There is no direct way to define the greatest lower bound of two
elements of $\s$. Indeed, two abstract values may represent the same
concrete value but without being comparable. Hence we only have a
join-semilattice structure. The meet operation $\sqcap_\s$ over
floating-point slopes is defined in
Equation~\eqref{eq:meet-operation}. It may require a conversion into
interval value. We consider that the result of the meet operation
introduces a new independent variable at index $\ell$. We denote by
$\sqsubset_\I$ the strict comparison of intervals and by $\bot_\s$ the
least element of $\s$. {\small
  \begin{multline}
    \label{eq:meet-operation}
    \forall (\interval{M}_g, \vinterval{S}_g), 
    (\interval{M}_h, \vinterval{S}_h) \in \s, \quad  
    \big(\interval{M}_g, \vinterval{S}_g\big) 
    \ \sqcap_\s\
    \big(\interval{M}_h, \vinterval{S}_h\big) = 
    \\
    \begin{cases}
      \bot_\s 
      & \text{if } \iota(\interval{M}_g, \vinterval{S}_g) 
      \sqcap_\I \iota(\interval{M}_h, \vinterval{S}_h) = \bot_\I
      \\
     (\interval{M}_g, \vinterval{S}_g) 
     & \text{if } \iota(\interval{M}_g, \vinterval{S}_g) 
     \sqsubset_\I \iota(\interval{M}_h, \vinterval{S}_h)
     \\
     (\interval{M}_h, \vinterval{S}_h) 
     & \text{if } \iota(\interval{M}_h, \vinterval{S}_h) 
     \sqsubset_\I \iota(\interval{M}_g, \vinterval{S}_g)
     \\
     \kappa\big(\iota(\interval{M}_h, \vinterval{S}_h) 
     \sqcap^\ell_\I \iota(\interval{M}_g, \vinterval{S}_g)\big) 
     & \text{otherwise}
    \end{cases}
  \end{multline}
}%

\subsubsection*{Note on the Widening Operator.}
\label{sec:acceleration-of-convergence}

In order to enforce the convergence of the fixpoint computation, we
can define a widening operation $\nabla_\s$ over floating-point slopes
values. An advantage of our domain is that we can straightforwardly
use the widening operations defined for the interval domain denoted by
$\nabla_\I$. We define the operator $\nabla_\s$ in
Equation~\eqref{eq:widening-operation} using the widening operator
between intervals. The notation $\dot{\nabla}_\I$ represents the
component-wise application of $\nabla_\I$ between the components of
the interval slopes vector. {\small
  \begin{multline}
    \label{eq:widening-operation}
    \forall (\interval{M}_g, \vinterval{S}_g), 
    (\interval{M}_h, \vinterval{S}_h) \in \s,\quad
    \big(\interval{M}_g, \vinterval{S}_g\big) \nabla_\s 
    \big(\interval{M}_h, \vinterval{S}_h\big) = 
    \big(\interval{M}, \vinterval{S}\big)
    \\
    \text{with}\quad
    \interval{M} = \interval{M}_g\ \nabla_\I\ \interval{M}_h 
    \quad\text{and}\quad \vinterval{S} = \vinterval{S}_g\
    \dot{\nabla}_\I\ \vinterval{S}_h
  \end{multline}
}%

\section{Analysis of Floating-Point Programs}
\label{sec:analysis-of-floating-point-programs}

The goal of the static analysis of floating-point programs using the
floating-point slopes domain is to give for each control point and for
each variable an over-approximation given by \textsf{FPS} of the
reachable set of floating-point numbers. An abstract environment
$\text{env}^\sharp$ associates to each variable $v\in \mathcal{V}$ a
value of $\s$. The set $\mathcal{V}$ is made of the sets $\varind$ and
$\vardep$ of independent and dependent variables.

The semantics of an assignment $\osem{v := e}$ in the abstract
environment $\text{env}^\sharp$ is the update of the value associated
to $v$ with the result of the evaluation of the arithmetic expression
$e$ using the arithmetic operations over \textsf{FPS} given in
Figure~\ref{fig:arithmetic-operation-floating-point-slopes}. As the
\textsf{FPS} domain is related to the interval domain we can
straightforwardly use the semantics of tests given in \cite{Granger92}
to refine the value of variables. Note that the semantics of tests is
related to the meet operation defined in
Equation~\eqref{eq:meet-operation} which may conserve some relations
between variables.

We define in Equation~\eqref{eq:concretization} the concretization
function $\gamma_\s$ between the join-semilattice $\langle \mathcal{V}
\rightarrow \s, \dot{\sqsubseteq}_\s \rangle$, with
$\dot{\sqsubseteq}_\s$ the point-wise lifting comparison, and the
complete lattice $\langle \wp(\mathcal{V} \rightarrow \F), \subseteq
\rangle$. 
{\small
\begin{equation}
  \label{eq:concretization}
  \gamma_\s \big( v \mapsto \left(\interval{M}, \vinterval{S}\right)\big) = 
  \bigcup_{\vscalar{u} \in \vinterval{V}_\varind} 
  \big\{ v \mapsto i \in \interval{I}: \interval{I} =
    \interval{M} +  \vinterval{S} 
    \cdot \left(\vscalar{u} - \fun{mid}\big(\vinterval{V}_\varind\big)\right)
  \big\}
\end{equation}
}In Theorem~\ref{thm:soudness}, we state the soundness of the
floating-point analysis using \texttt{FPS} domain with respect to the
concrete floating-point semantics. The later is based on the concrete
semantics of floating-point expressions $\llbracket e \rrbracket$, see
\cite{Min04} for its definition.

\begin{theorem}
  \label{thm:soudness}
  If the set of concrete environments $\text{env}$ is contained in the
  abstract environment $\text{env}^\sharp$ then we have for all
  instruction $i$ representing either an assignment or a test:{\small
  \begin{displaymath}
    \llbracket i \rrbracket(\text{env}) 
    \subseteq 
    \gamma_\s\left(\llbracket i
      \rrbracket^\sharp(\text{env}^\sharp)\right)
    \enspace.
  \end{displaymath}}
\end{theorem}

\section{Case Studies}
\label{sec:case-studies}

In this section, we present experimental results of the static
analysis of numerical programs using our floating-point slope
domain. We based our examples on Matlab/Simulink models which are
block-diagrams. We present as examples a second order linear filter
and a square root computation with a Newton method.

We first give a quick view of Matlab/Simulink models. In a
block-diagram, each node represents an operation and each wire
represents a value evolving during time. We consider a few operations
such that arithmetic operations, gain operation that is multiplication
by a constant, conditional statement (called
\textit{switch}\footnote{This operation is equivalent to the
  conditional expression: $\text{if } p_c(e_0) \text{ then } e_1
  \text{ else } e_2$. The predicate $p_c$ has the form $e_0 \diamond
  c$ where $c$ is a given constant and $\diamond \in \{\ge, >, \neq\}$.}
in Simulink), and \textit{unit delay} block represented by
$\frac{1}{\text{z}}$ which acts as a memory. We can hence write
discrete-time models thanks to finite difference equations, see
\cite{CM09b} for further details.

The semantics of Simulink models is based on finite-time execution.
In other words, a Simulink model is implicitly embedded in a
\textit{simulation loop} modelling the temporal evolution starting
from $t=0$ to a given final time $t_{\text{end}}$. The body of this
loop follows three steps: \textit{i)}~evaluating the inputs,
\textit{ii)}~computing the outputs, \textit{iii)}~updating the state
variables \ie values of the unit delay blocks. The static analysis of
Simulink models transforms the simulation loop into a fixpoint
computation. In its simple form, see \cite{CM09b} for further details,
we add an extra time instant to collect all the behaviors from
$t_{\text{end}}$ to $t=+\infty$.

\subsubsection*{Linear Filter.}
\label{sec:linear-filter}

\begin{figure}[t]
  \centering
  \subfigure[Simulink model]{
    \includegraphics[scale=0.25,angle=-90]{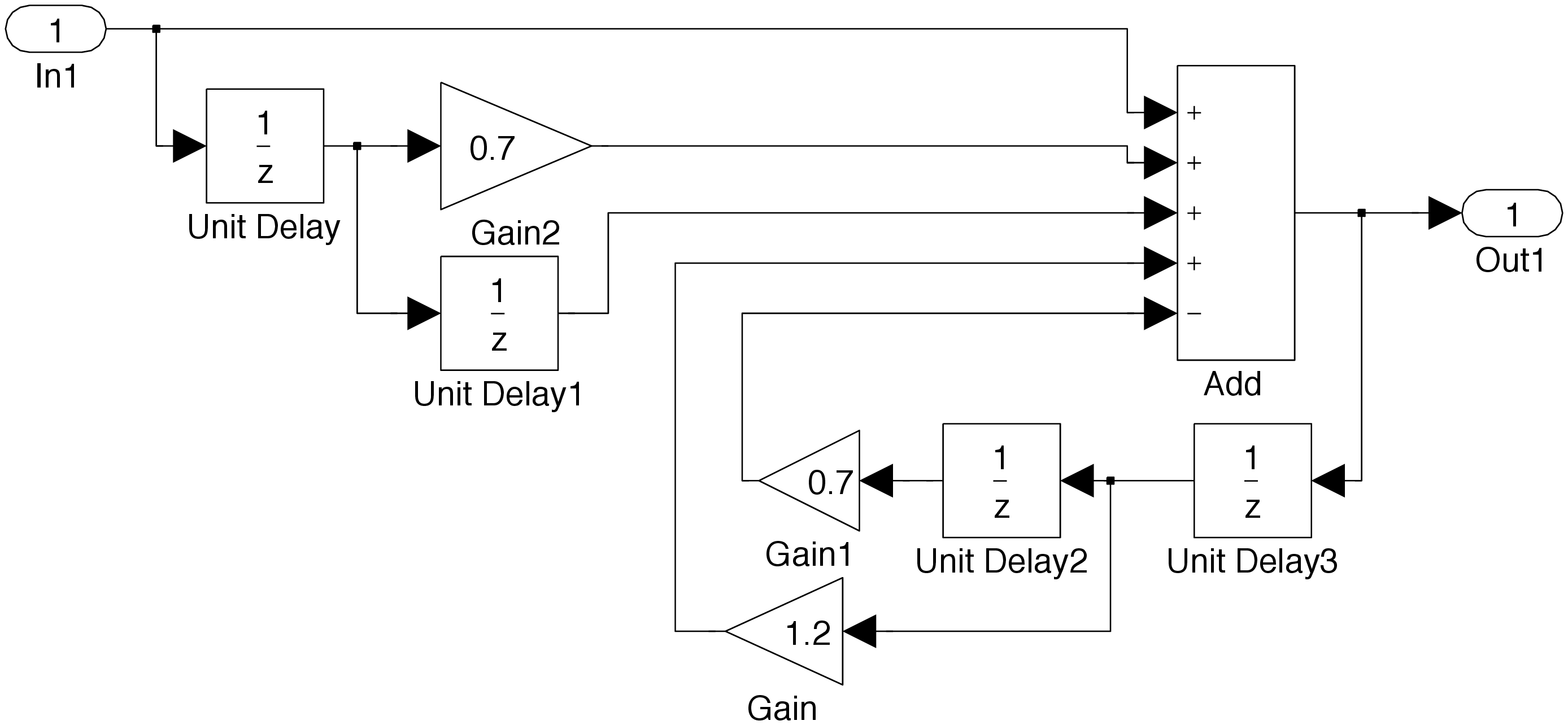}
    \label{fig:bd-linear-filter}
  }
  \subfigure[Temporal evolution of the output]{
    \includegraphics[scale=0.2025,angle=-90]{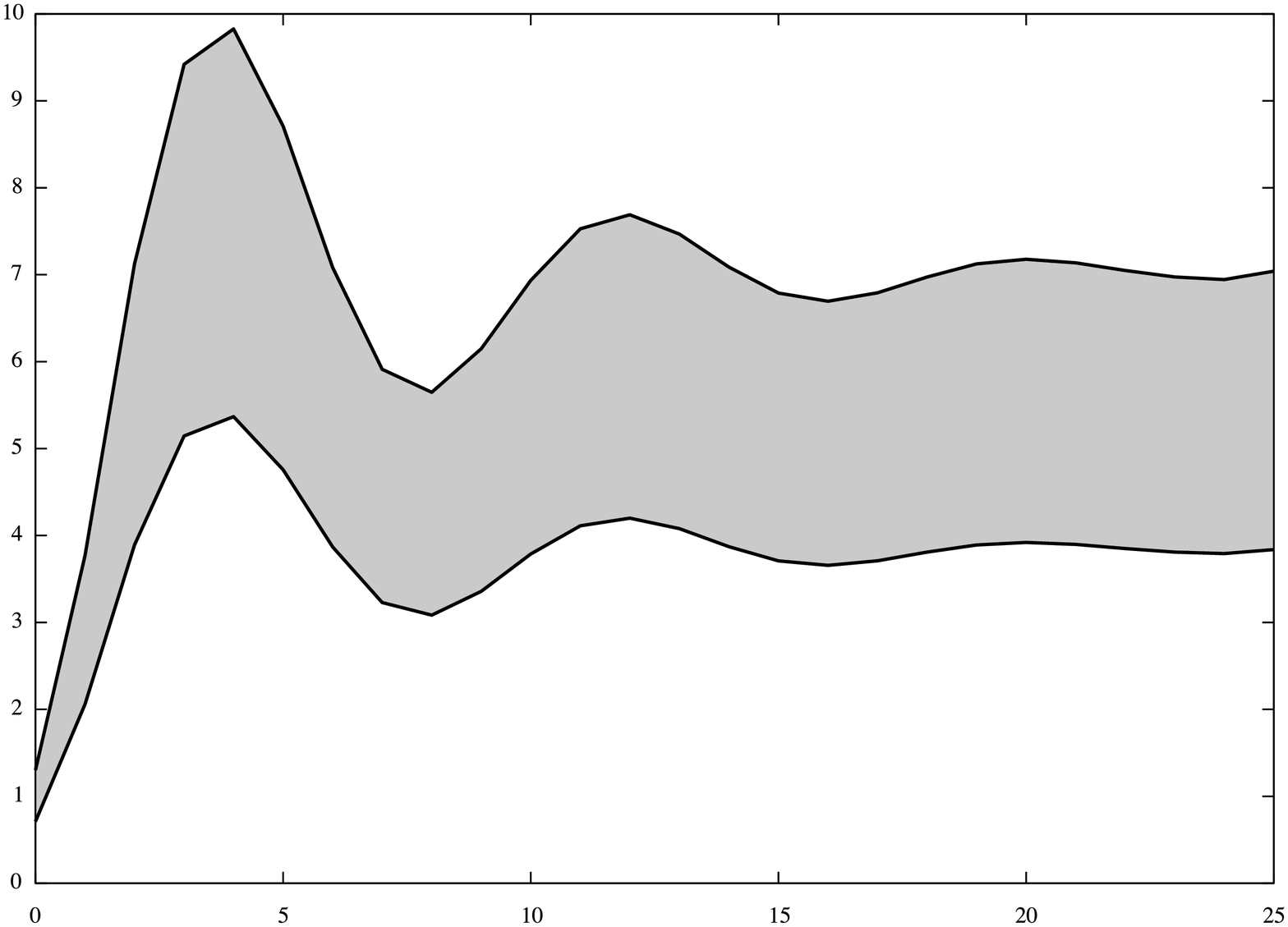}
    \label{fig:output-linear-filter}
  }
  \caption{Second order linear filter}
  \label{fig:linear-filter}
\end{figure}

We applied the floating-point slope domain on a second order linear
filter defined by:
\begin{math}
  y_n = x_n + 0.7 x_{n-1} + x_{n-2} + 1.2 y_{n-1} - 0.7 y_{n-2}\enspace.
\end{math}

The block-diagrams of this filter is given in
Figure~\ref{fig:bd-linear-filter}. We consider a simulation time of
$25$ seconds that is we unfold the simulation loop $25$ times before
making unions. The input belongs into the interval $[0.71, 1.35]$. The
output of the filter is given in
Figure~\ref{fig:output-linear-filter}. We consider, in this example,
that $\varind$ contains the input and the four unit delay blocks that
is there are five independent variables. The gray area represents all
the possible trajectories of the output corresponding of the set of
inputs. Hence we can bound the output, without using the widening
operator, by the interval $[0.7099, 9.8269]$.

\subsubsection*{Newton Method.}
\label{sec:nonlinear-algorithms}

\begin{figure}[t]
  \centering
  \subfigure[Main model]{
    \includegraphics[scale=0.25,angle=-90]{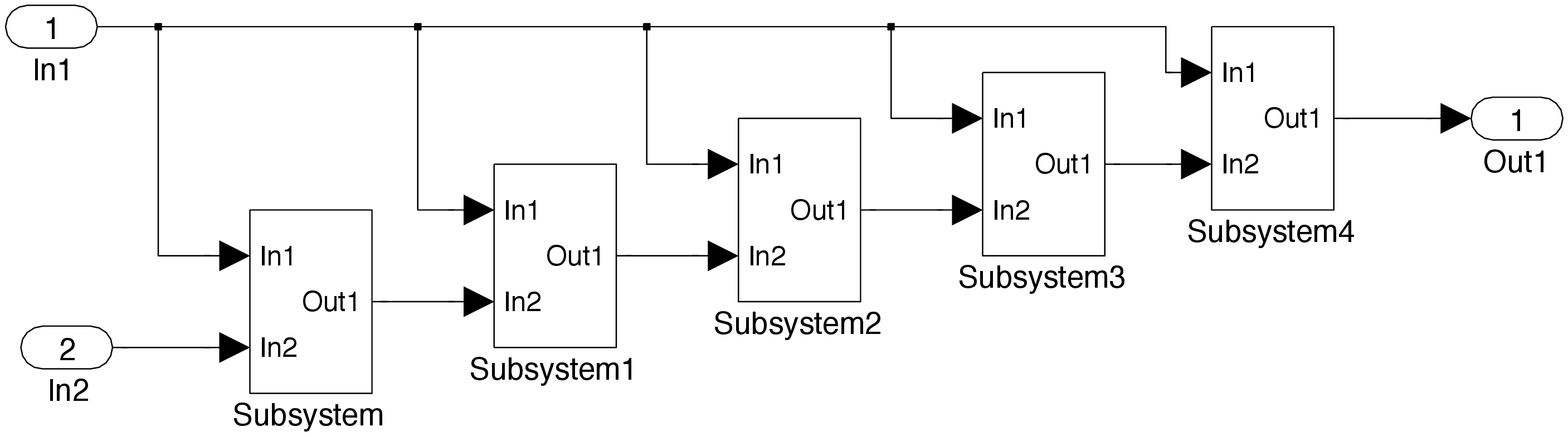}
    \label{fig:newton-model}
  }
  \subfigure[Content of a subsystem]{
    \includegraphics[scale=0.25,angle=-90]{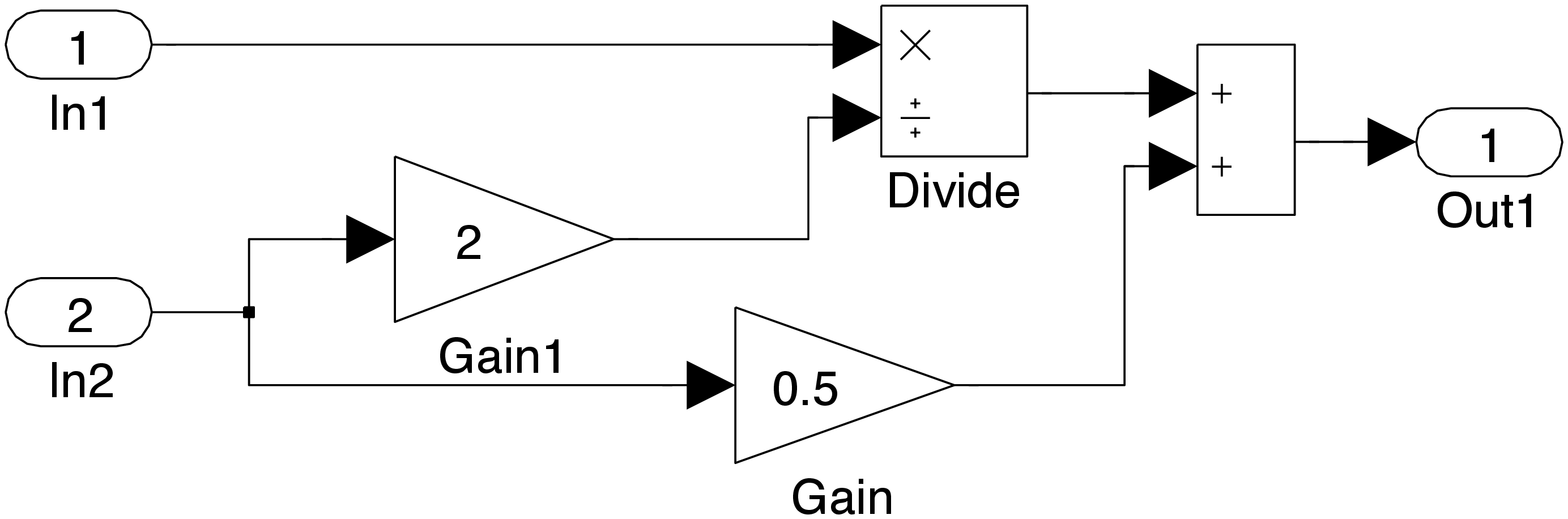}
    \label{fig:content-newton}
  }
  \caption{Simulink model of the square root computation}
  \label{fig:simulink-newton}
\end{figure}

We applied our domain on a Newton algorithm which computes the square
root of a number $a$ using the following iterative sequence:
\begin{math}
  x_{n+1} = \frac{x_n}{2} + \frac{a}{2x_n} \enspace.
\end{math}

We want to compute $x_5$ that is we consider the result of the Newton
method after five iterations. The Simulink model is given in
Figure~\ref{fig:newton-model} and in Figure~\ref{fig:content-newton},
we give the model associated to one iteration of the algorithm. In
this case, the set $\varind$ is only made of one element. For the
interval input $[4, 8]$ with the initial value equals to $2$, we have
the result $[1.8547, 3.0442]$.

\section{Related Work}
\label{sec:related-work}

Numerical domains have been intensively studied. A large part of
numerical domains concern the polyhedral representation of sets. For
example, we have the domain of polyhedron \cite{CH78} and the variants
\cite{SKH03,mine:hosc06,SCSM06,PH07,CC07,LF08,LL09,CMWC09,CMWC10}. We
also have the numerical domains based on affine relations between
variables \cite{Karr76,GGP09} or the domain of linear congruences
\cite{Granger91}. In general, all these domains are based on
arithmetic with "good" properties such that rational numbers or real
numbers. A notable exception is the floating-point versions of the
octagon domain \cite{Min04} and of the domain of polyhedron
\cite{CMC09}. These domains give a sound over-approximation of the
floating-point behaviors but they are not empowered to model the
behaviors of floating-point arithmetic as we do.

Our \textsf{FPS} domain is more general than numerical abstract
domains made for a special purpose. For example, we have the domain
for linear filters \cite{Fer04} or for the numerical precision
\cite{GP06} which provide excellent results. Nevertheless as we showed
in Section~\ref{sec:case-studies}, we can apply this domain in various
situations without losing too much precision.

\section{Conclusion}
\label{sec:conclusion}

We presented a new partially relational abstract numerical domain
called \textsf{FPS} dedicated to floating-point variables. It is based
on Krawczyk and Neumaier's work \cite{KN85} on interval expansion of
rational function using interval slopes. This domain is able to mimic
the behaviors of the floating-point arithmetic such that the
\textit{absorption} phenomenon. We also presented experimental results
showing the practical use of this domain in various contexts.

We want to pursue the work on the \textsf{FPS} domain by refining the
the meet operation in order to keep relations between
variables. Moreover we would like to model more closely the behaviors
of floating point arithmetic, for example by taking into account the
hardware instructions \cite[Sect.~3]{Mon05}.

As an other future work, we want to apply \textsf{FPS} domain for the
analyses of the numerical precision by combining the \textsf{FPS}
domain and domains defined in \cite{Mar04b,CM09a}. An interesting
direction should be to make an analysis of the numerical precision by
comparing results of the \textsf{FPS} domain and results coming from
the other numerical domain which bound the exact mathematical
behaviors such that \cite{CMC09}. Hence we can avoid the manipulation
of complex abstract values to represent rounding errors such as in
\cite{Mar04b,GP06,CM09a}.

\subsubsection*{Acknowledgements.}
The author deeply thanks O.~Bouissou, S.~Graillat, T.~Hilaire,
D.~Mass\'e and M.~Martel for their useful comments on the earlier
versions of this article. He is also very grateful to anonymous
referees who helped improving this work.

\bibliographystyle{plain}
\bibliography{biblio}

\end{document}